\documentclass[sn-mathphys-num]{sn-jnl}

\usepackage{graphicx}%
\usepackage{multirow}%
\usepackage{amsmath,amssymb,amsfonts}%
\usepackage{amsthm}%
\usepackage{mathrsfs}%
\usepackage[title]{appendix}%
\usepackage{xcolor}%
\usepackage{textcomp}%
\usepackage{manyfoot}%
\usepackage{booktabs}%
\usepackage{algorithm}%
\usepackage{algorithmicx}%
\usepackage{algpseudocode}%
\usepackage{listings}%

\theoremstyle{thmstyleone}%
\theoremstyle{thmstyletwo}%

\theoremstyle{thmstylethree}%
\def\prl{Phys.~Rev.~Lett.} 
\def\prc{Phys.~Rev.~C} 
\def\prd{Phys.~Rev.~D} 
\def\aap{Astron. Astrophys.}   
\def\physrep{Phys.~Rep.}   
\def\nphysa{Nucl.~Phys.~A}   
\def\mnras{Mon. Not. R. Astron. Soc.}             

\begin{document}

\title[Filling fractions for the formation of nuclear pasta in neutron stars]{Filling fractions for the formation of nuclear pasta in neutron stars: semiclassical vs liquid-drop predictions}


\author*[1]{\fnm{Nikolai N.} \sur{Shchechilin}}\email{nikolai.shchechilin@ulb.be}

\author[1]{\fnm{Nicolas} \sur{Chamel}}\email{nicolas.chamel@ulb.be}
\equalcont{These authors contributed equally to this work.}

\author[2]{\fnm{Andrey I.} \sur{Chugunov}}\email{andr.astro@mail.ioffe.ru}
\equalcont{These authors contributed equally to this work.}

\affil[1]{\orgdiv{Institut d'Astronomie et d'Astrophysique}, \orgname{Universit\'e Libre de Bruxelles}, \orgaddress{\street{CP226}, \city{Brussels}, \postcode{1050}, \country{Belgium}}}

\affil[2]{\orgname{Ioffe Institute}, \orgaddress{\street{Politeknicheskaya 26}, \city{Saint Petersburg}, \postcode{194021}, \country{Russia}}}


\abstract{Historically, a sequence of nuclear pasta shapes was predicted to appear in the deepest region of the inner 
crust of a neutron star within the compressible liquid-drop picture, when the filling fraction $u$ exceeds some 
threshold values. However, later calculations showed that these values depend on the details of the liquid-drop 
model. Here we investigate the existence of pasta 
in neutron stars within the semiclassical extended Thomas-Fermi approach using various generalized 
Skyrme functionals. The filling fractions for the different transitions are found to be quasi-universal, unlike 
the pasta density ranges governed by the symmetry energy at relevant densities. In particular, pasta emerge at $u_\mathrm{sp}\approx0.13-0.15$. By applying a simplified 
stability criterion within the liquid-drop framework, we show that these values of $u_\mathrm{sp}$ can be explained by 
the nuclear curvature correction. In this way, the abundance of pasta can be easily estimated. This criterion can also 
be used to optimize the search of pasta within the more realistic extended Thomas-Fermi approach.  
}




\maketitle

\section{Introduction}\label{sec1}

The inner crust of a neutron star consists of nuclear clusters immersed in a sea of free nucleons and electrons.   
It spans more than two orders of magnitude in density up to half of the nuclear saturation density $0.5 n_\mathrm{sat}$ 
where a phase transition to the homogeneous matter of the core takes place.
For non-accreted neutron stars under the cold-catalyzed matter hypothesis, only one type of nuclear clusters is present at the given mean baryon density $\bar{n}$. At relatively low $\bar{n}$, their filling fraction $u$ is much smaller than 1, and their shape is expected to be quasi-spherical (such clusters are often called ``nuclei'' even though they could not exist in a vacuum due to their extreme neutron richness). Meanwhile, the optimal shape can change drastically as $u$ approaches 1. Already in one of the pioneering works on the neutron star crust structure within the compressible liquid drop model (CLDM), Baym, Bethe and Pethick \cite{BBP1971} predicted that beyond $u=0.5$ the nuclei can ``turn inside out'' forming a bubble phase. Later, two groups, Ravenhall, Pethick, and Wilson \cite{Ravenhall_ea83} and Hashimoto, Seki, and Yamada \cite{Hashimoto+84} independently realized that the transition from spherical clusters proceeds through several shapes. The transformations of spheres into rods, plates, tubes, and bubbles became the traditional sequence, often referred to as {\it pasta} due to the resemblance of the non-spherical shapes to spaghetti, lasagna, and bucatini respectively (bubbles are usually referred to as ``Swiss cheese'' to follow the culinary analogy). 
While in Ref.~\cite{Ravenhall_ea83} the simplified model of a particular nuclear Hamiltonian (Skyrme 1') was used, the conclusions of Ref.~\cite{Hashimoto+84} were more general. Within the liquid-drop approach (without curvature contribution), the authors of Ref.~\cite{Hashimoto+84} proved that the same pasta sequence appears for specific filling fractions regardless of nuclear interaction. 
With the help of a refined criterion, Oyamatsu, Hashimoto and Yamada \cite{Oyamatsu_ea84} established the pasta transitions from spheres to spaghetti, lasagna, bucatini, and Swiss cheese at the threshold filling fractions $u_\mathrm{t}=[0.19,0.35,0.65,0.81]$ respectively (note that clusters disappear at filling fractions $u_{\rm t,cl}$, while their inverted configurations appear at $1-u_{\rm t,cl}$).

In the review by Pethick and Ravenhall \cite{PethickRavenhall1995}, the Bohr-Wheeler condition for fission of isolated nuclei was invoked to show that a spherical nucleus is unstable against quadrupole deformations when it fills 1/8 of the Wigner-Seitz (WS) cell at variance with the predictions of Oyamatsu, Hashimoto and Yamada. The value $1/8$ has often been adopted as the filling fraction threshold $u_\mathrm{sp}$ for the transition from spheres to pasta. However, Zemlyakov and Chugunov \cite{Zemlyakov+22} have recently shown that for the catalyzed matter the neutralizing background gas of electrons suppresses the fission instability and the Bohr-Wheeler criterion is therefore not applicable. Their thermodynamically consistent CLDM with adsorbed neutrons (but without curvature effects) predicts a transition around $u_\mathrm{sp}\approx0.21$ in line with the work \cite{Oyamatsu_ea84} (see also, Ref.~\cite{Lim&Holt17}). 

In several studies~\cite{Pethick_ea83, Nakazato_ea11,Shchechilin+etf22,DinhThi+21b} it was argued that the curvature corrections can decrease the density (therefore the filling fraction) for the pasta formation. Indeed, in the systematic calculations of Ref.~\cite{Newton_ea13_Survey}  including curvature corrections, the transition from spheres to pasta was found to lie in the range $u_\mathrm{sp}\approx 0.08-0.14$, close to the value of $1/8=0.125$ from the Bohr–Wheeler fission instability criterion. On the other hand, in the CLDM of Refs.~\cite{DHM00,DH00,DH01} including curvature correction, spherical nuclei 
survived
up to the crust-core transition despite the filling fraction reaching $\approx 0.28$. 
In the more realistic Thomas-Fermi approach, the filling fraction at the transition to pasta was found to be close to 1/8~\cite{Oyamatsu&Iida07,Bao&Shen15,Okamoto+13}. 
This conclusion still holds when the Thomas-Fermi method with generalized Skyrme interactions is extended up to fourth order~\cite{Pearson+20}. 

Here, we investigate further the formation of nuclear pasta within the semiclassical extended Thomas-Fermi (ETF) and CLDM approaches. 
We first study in Sec.~\ref{sec2} the sensitivity of the pasta formation within the ETF method considering the set of generalized Skyrme interactions BSk22, BSk24, BSk25~\cite{Goriely_ea_Bsk22-26}. These functionals were precision-fitted to essentially all nuclear masses with different values of the symmetry energy coefficient at saturation $J=32,30,29$ MeV, respectively, and were simultaneously constrained to reproduce a realistic neutron-matter equation of state. In Sec.~\ref{sec3}, we then revisit the role of curvature corrections in CLDM based on the same interactions. For comparison, we also consider the popular SLy4 interaction~\cite{Chabanat_ea98_SLY4} used in Refs.~\cite{DHM00,DH00,DH01}.  Our conclusions are outlined in Sec.~\ref{sec4}.

\section{Extended Thomas-Fermi method}\label{sec2}

The approach we follow is described in detail in Ref.~\cite{Shchechilin+sym23}, where we investigated the role of symmetry energy in the abundance of pasta structures. However, we implement some modifications here. First, we employ the soft-damping parametrization for the nucleon density profiles $n_q (\pmb{r})$ with $q=n,p$ for neutrons, protons introduced in Ref.~\cite{ShchechilinParam+23}. This new parametrization was shown to be more realistic for the dense crustal layers where pasta may form than the strong-damping variant adopted before. To compare with CLDM, we do not include the microscopic proton shell and pairing corrections. Finally, since we focus here on the densest part of the crust where nucleons are more smoothly distributed, we truncate the semiclassical expansion of kinetic energy and spin current densities in terms of the local nucleon number densities and their gradients at the second order~\cite{Brack_ea85}. Then the total energy of the WS cell becomes a functional of the form $E_\mathrm{ETF}[n_q (\pmb{r}),\nabla n_q(\pmb{r})]$. This total energy divided by the number of nucleons in the cell is minimized at the given mean nucleon number density $\bar{n}$ considering spherical, cylindrical and platelike WS cells for spheres/Swiss cheese, spaghetti/bucatini and lasagna respectively.
This allows us to compare the resulting energies per particle among 5 shapes (spheres/spaghetti/lasagna with central density accumulation and bucatini/Swiss cheese with central density depletion) and to obtain the equilibrium configuration. 

\begin{figure}[h]
\centering
\includegraphics[width=0.7\textwidth]{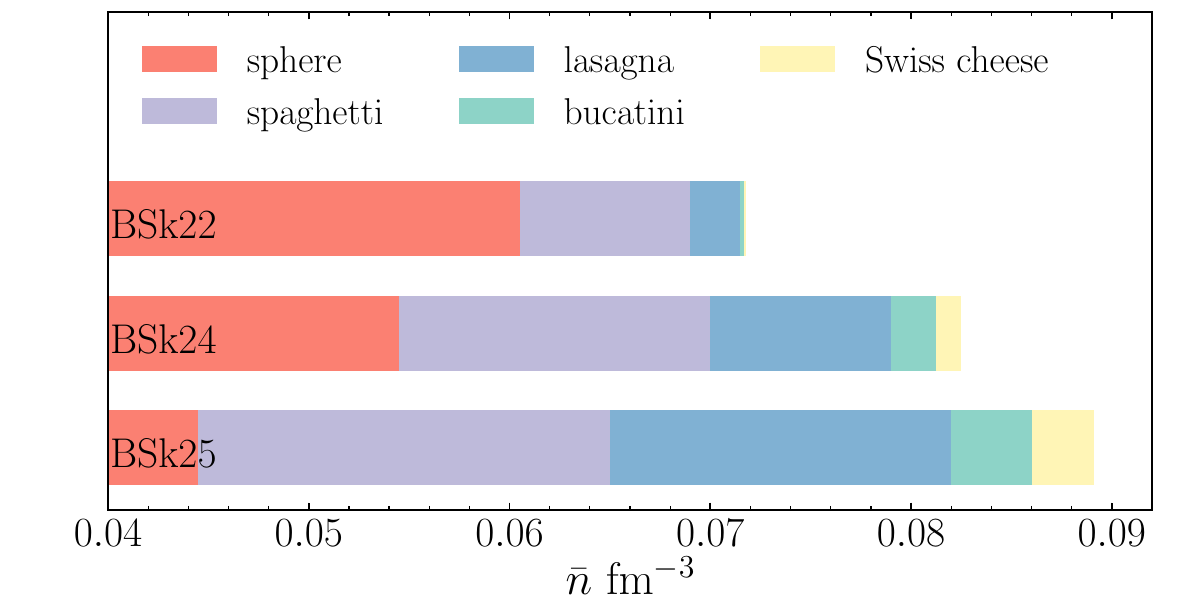}
\caption{Pasta sequences with BSk22, BSk24 and BSk25 Skyrme parametrizations.}\label{fig:n_sequence}
\end{figure}

In this way, we compute the pasta sequences for the BSk22, BSk24, and BSk25 functionals~\cite{Goriely_ea_Bsk22-26} that differ in the behavior of the symmetry energy $S(n)$ with density: 
for the density region considered in this work, $S(n)$ systematically grows from BSk22 to BSk25 
(see Fig.\ 3 in Ref.~\cite{Pearson_ea18_bsk22-26}).\footnote{Note, that for the applied BSk functionals a higher symmetry-energy coefficient $J$ at saturation leads to lower symmetry energy for inner crust densities due to the tight constraints around $n\approx0.11$ fm$^{-3}$ imposed by the precision fitting to the experimental atomic masses.}  
The results are presented in Fig.~\ref{fig:n_sequence}. As in Ref.~\cite{Shchechilin+sym23}, we observe an expansion of the density region filled by pasta with an increase of $S(n)$ at the relevant densities. However, with the new more realistic ETF profile parametrizations, lasagna covers a larger density range for all functionals, as predicted in Ref.~\cite{ShchechilinParam+23}.   

In the spirit of the liquid-drop picture, we define the size of proton clusters such that the uniform densities $n_{{\rm o}p}$ outside and $n_{{\rm i}p}$ inside are the same as the density at the WS cell border $n_{{\rm B}p}$ and the central density $n_{\Lambda p}+n_{{\rm B}p}$ of the parametrized ETF proton density distribution $n_p(\pmb{r})$ respectively, keeping the overall number of protons fixed. The filling fraction is then given by 
\begin{align}\label{eq:u_ETF_cluster}
   u= \dfrac{1}{V_{\rm cell} n_{\Lambda p}} \int_{\rm cell} [n_p(\pmb{r})-n_{{\rm B}p}] d^3\pmb{r}\quad ,
\end{align}
where $V_{\rm cell}$ denotes the WS cell volume.\footnote{For spaghetti/bucatini and lasagna, the WS cells consist of infinitely long cylinders and slabs of infinite extent, respectively. The integrations, however, are performed for a finite cell of unit length and unit area respectively} Likewise, for the inverse phases (holes) we obtain
\begin{align}\label{eq:u_ETF_hole}
   u= 1- \dfrac{1}{V_{\rm cell} n_{\Lambda p}} \int_{\rm cell} [n_p(\pmb{r})-n_{{\rm B}p}] d^3\pmb{r} \quad.  
\end{align} 

\begin{figure}[h]
\centering
\includegraphics[width=0.65\textwidth]{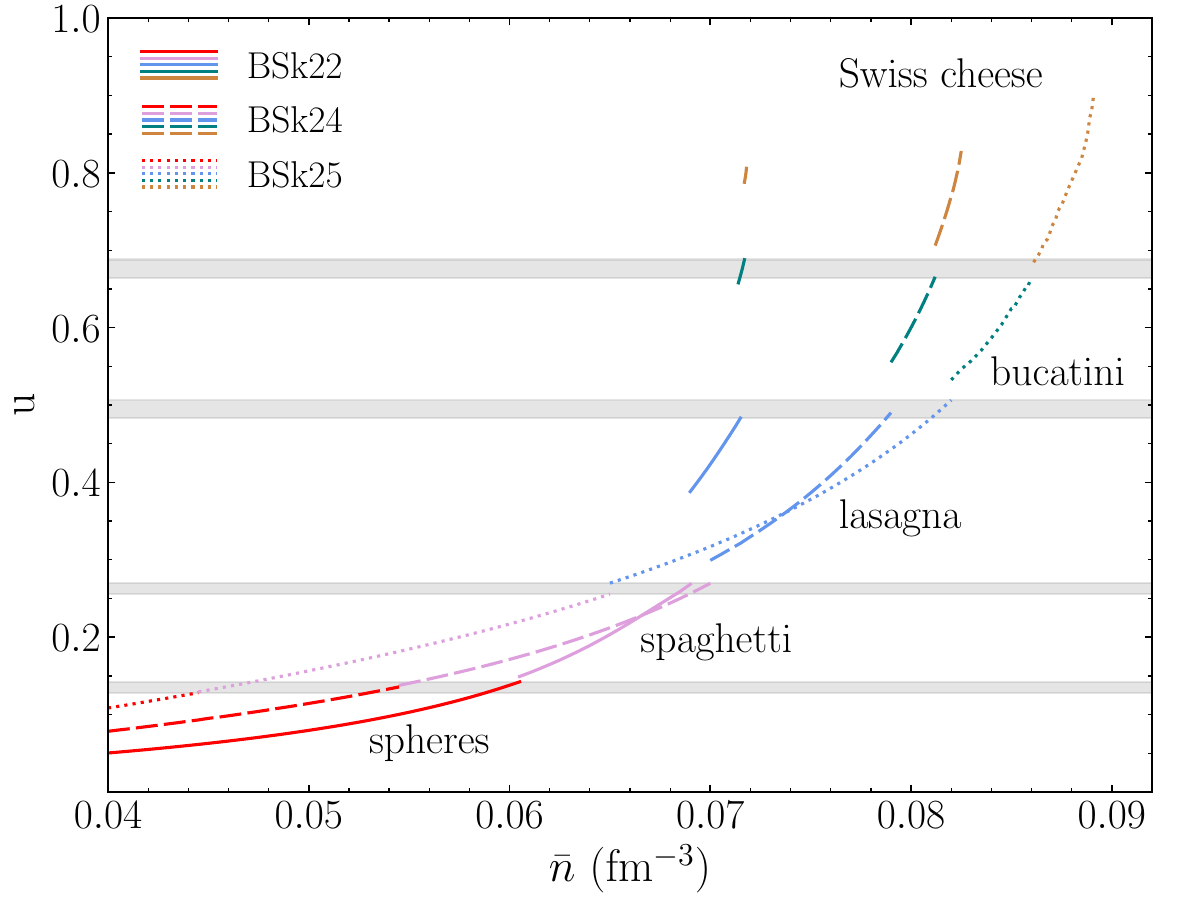}
\caption{Filling fractions and corresponding pasta configurations 
versus mean baryon density with BSk22 (solid lines), BSk24 (dashed lines) and BSk25 (dotted lines) generalized Skyrme parametrizations. The grey horizontal bands mark the upper limit of existence of each shape.}\label{fig:u_vs_n}
\end{figure}

Fig.~\ref{fig:u_vs_n} shows the evolution of $u$ and the corresponding equilibrium pasta configuration with the mean baryon density. It is interesting to note, that while the transition densities between different shapes vary significantly for the applied BSk functionals, the deviations of the corresponding filling fractions $u_\mathrm{t}$ are very small. In particular, the onset of pasta is found for $u_\mathrm{sp}\approx0.13-0.14$. Spaghetti, which are the first to appear, fill slightly more space for the same baryon density thus leading to a discontinuous change in $u$. 
When the filling fraction reaches $u_{\rm t}\approx0.26-0.27$, spaghetti transform to lasagna. The latter disappear at $u_{\rm t}\approx0.48-0.51$ and are replaced by bucatini up to $u_{\rm t}\approx0.66-0.69$. The densest layers of the crust consist of 
Swiss cheese. The quasi-universal $u_\mathrm{t}$ can serve as good indicators for the optimization of the ETF calculations in the search for the equilibrium pasta phase. 

\begin{figure}[h]
\centering
\includegraphics[width=0.65\textwidth]{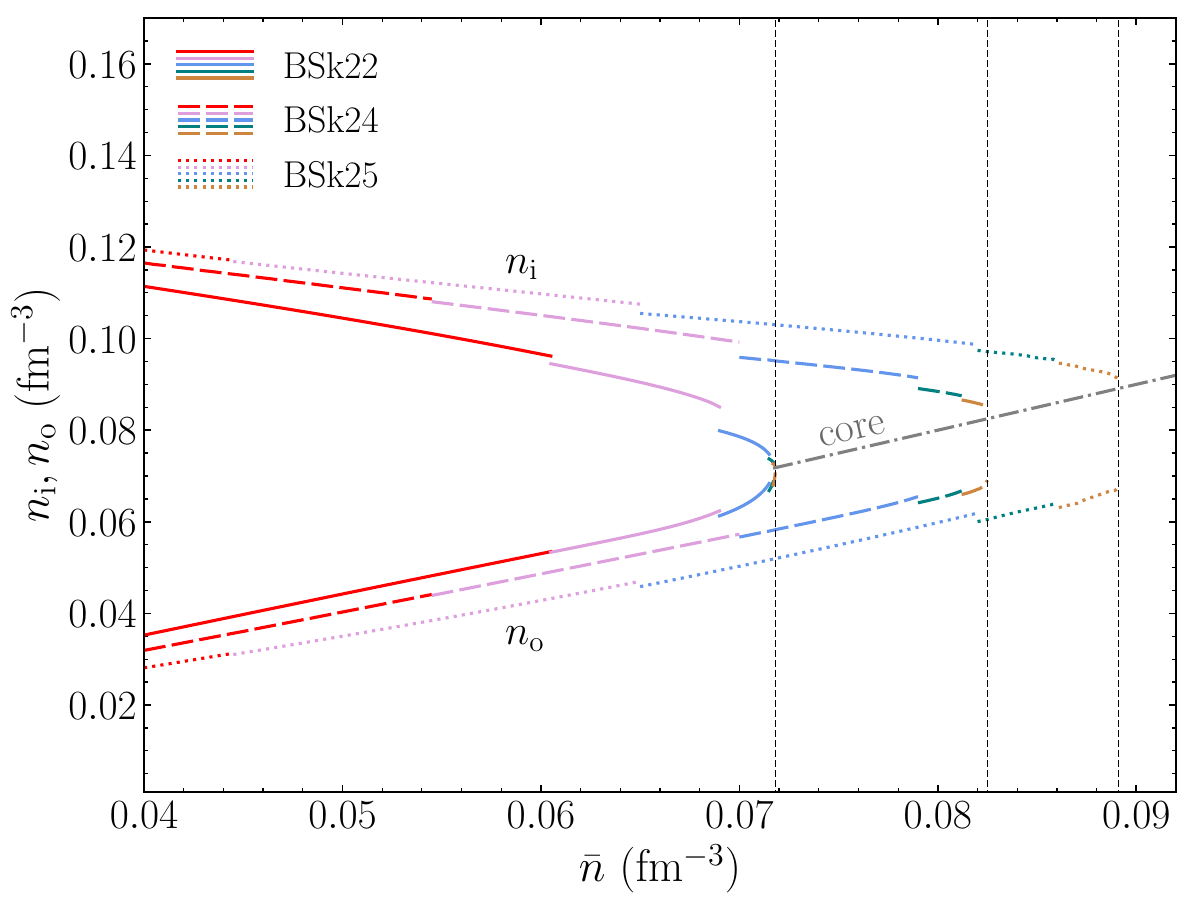}
\caption{The same is in Fig.~\ref{fig:u_vs_n} but for the densities $n_\mathrm{i}$ inside and $n_\mathrm{o}$ outside clusters. Black vertical lines cut off the homogeneous matter of the core for each functional. The corresponding density in the core is denoted by the dash-dotted grey line.}\label{fig:n_io_vs_n}
\end{figure}

Looking at Fig.~\ref{fig:u_vs_n}, we see that even though the symmetry energy has almost no impact on
$u_\mathrm{t}$, it governs the density dependence of $u(\bar{n})$.
This can be understood from the approximate formula
\begin{align}\label{eq:u_approx}
    u\approx \frac{\bar{n}-n_\mathrm{o}}{n_\mathrm{i}-n_\mathrm{o}} \quad .
\end{align}
Indeed, the density $n_\mathrm{i}$ ($n_\mathrm{o}$) inside (outside) clusters decreases (increases) when 
varying the functional from BSk25 to BSk22 corresponding to lowering $S(n)$ at the relevant densities (therefore proton fraction), 
as shown in Fig.~\ref{fig:n_io_vs_n}. Such a correlation is in agreement with the TF calculations of Refs.~\cite{Bao&Shen15,Oyamatsu&Iida07}. At relatively low densities, $n_\mathrm{o}\ll n_\mathrm{i}$ so that 
$u\approx (\bar{n}-n_\mathrm{o})/n_\mathrm{i}$. Then the relative variation of $u$ with the change of the applied functional is $\delta u/u\approx - \delta n_\mathrm{i}/n_\mathrm{i} - \delta n_\mathrm{o}/(\bar{n}-n_\mathrm{o})$. From Fig.~\ref{fig:n_io_vs_n}, we see 
that $\delta n_\mathrm{i}/n_\mathrm{i}$ is negligible in comparison with $\delta n_\mathrm{o}/(\bar{n}-n_\mathrm{o})$.
Then the decrease of $n_\mathrm{o}$ from BSk22 to BSk25 leads to an increase of $u$.
Therefore $u_\mathrm{sp}$ is reached at a density that is lower for BSk25 than for BSk22. At higher densities, $n_\mathrm{i}-n_\mathrm{o}$ decreases faster with $\bar n$ for BSk22 than for BSk25, therefore the slope of $u(\bar n)$ is higher, as demonstrated in Fig.~\ref{fig:u_vs_n}. This reduces the density range of each pasta phase and the crust dissolves at a lower density (see Fig.~\ref{fig:n_sequence}). { The latter fact points toward a correlation of the crust-core transition density with the symmetry energy at the relevant densities (for the functionals used here, this implies an anti-correlation with the symmetry-energy slope $L=3 n_\mathrm{sat} S'(n_\mathrm{sat})$ at saturation density, the prime denoting the derivative with respect to $n$, in agreement with Refs.~\cite{Ducoin+10_Symm,Gonzalez-Boquera_ea19}).}

Our calculations reveal a rather discontinuous transition from the clustered state with $u\lesssim0.9$ and $n_\mathrm{i}\neq n_\mathrm{o}$ to the homogeneous matter of the core, as depicted in Figs.~\ref{fig:u_vs_n},~\ref{fig:n_io_vs_n}.
This suggests that 
the transition is of first order. It is worth noting that the predicted crust-core transition agrees well with the instability of the core matter with respect to the formation of finite-size inhomogeneities
\cite{Pearson_ea18_bsk22-26},
i.e., the crust-core transition occurs at slightly larger densities where the core is stable. However, we should emphasize that the numerical determination of the equilibrium configuration 
becomes more and more delicate as the crust-core boundary is approached; moreover, the WS approximation and the choice of the profile parametrization can also affect the calculated values of the filling fraction at the very bottom of the crust.

\section{Compressible liquid drop model}\label{sec3}

\subsection{Key equations including neutron skin and curvature}\label{sec31}

Having identified the pasta transitions in the ETF calculations, we now examine more closely the role of surface and Coulomb effects on the determination of the equilibrium shape within the CLDM. For each pasta shape, we introduce an elementary electrically charged neutral cell of volume $a^3$ such that it contains
one cluster or hole. 
For spaghetti and bucatini, the faces of the cell perpendicular to the symmetry axis have area $a^2$. For lasagna, $a$ represents the shortest distance between slabs. 
The energy of each cell,  
\begin{align}\label{eq:CLDM-energy}
   E_\mathrm{CLD} = E_\mathrm{bulk} + E_\mathrm{s,pl} + E_\mathrm{s,curv} +E_\mathrm{Coul} \quad,
\end{align}
is divided into bulk, plane surface, curvature correction and Coulomb terms respectively. 
The equilibrium configuration at given $\bar{n}$ is then determined by comparing the energy densities $E_\mathrm{CLD}/a^3$ of all different shapes.

\begin{figure}[h]
\centering
\includegraphics[width=\textwidth]{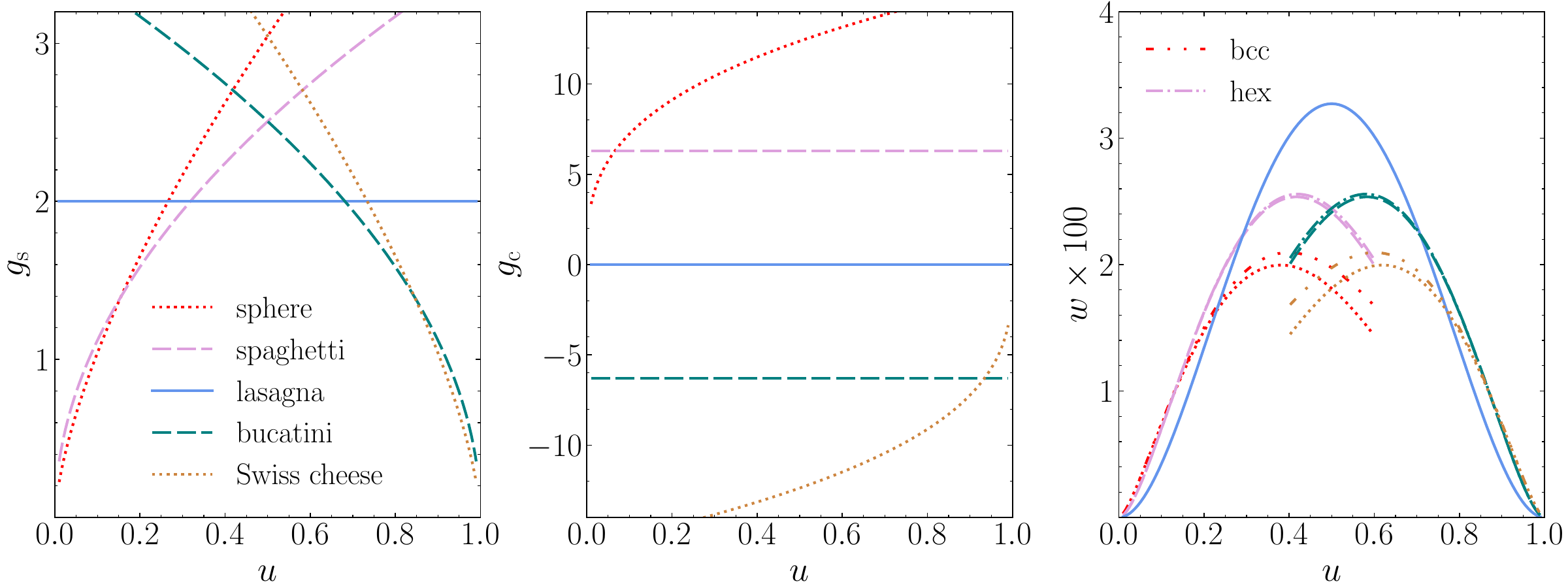}
\caption{Relative surface area $g_\mathrm{s}$ (left panel) and curvature correction function $g_\mathrm{c}$ (central panel) from Eq.~\eqref{eq:surf} for different pasta phases. Dimensionless Coulomb energy function $w$ (right panel) from Eq.~\eqref{eq:Coul} within the WS approximation (the same legend as in the left panels) as well as for body-centered cubic and hexagonal lattices (legend from the right panel).  }\label{fig:gs_gc_wc}
\end{figure}

The first term in Eq.~\eqref{eq:CLDM-energy}, the bulk energy scales as $E_\mathrm{bulk}=\mathcal{E}_\mathrm{bulk}(n_\mathrm{i},y_p,n_{\mathrm{o}n},n_{\mathrm{o}p},u)a^3$ where $\mathcal{E}_\mathrm{bulk}(n_\mathrm{i},y_p,n_{\mathrm{o}n},n_{\mathrm{o}p},u)$ is a function of the total nucleon density $n_\mathrm{i}$ and proton fraction $y_p$ inside the cluster (outside the hole), the neutron and proton densities $n_{\mathrm{o}n},n_{\mathrm{o}p}$ outside the cluster (inside the hole), and the filling fraction $u$, but does not depend explicitly on the shape of the cluster or hole. The surface terms in the thermodynamically consistent approach \cite{Centelles_ea98,DHM00,GC20_DiffEq,SC20_JPCS} should include $N_\mathrm{s}$ nucleons (in the case of very neutron-rich matter of neutron stars - neutrons) adsorbed on the surface. Neglecting the dynamical curvature corrections  \cite{DHM00,Centelles_ea98}, these terms can be written as
\begin{align}\label{eq:surf}
     E_\mathrm{s,pl} + E_\mathrm{s,curv} = \sigma_\mathrm{s} g_\mathrm{s}(u) a^2 + \sigma_\mathrm{c} g_\mathrm{c}(u) a + N_\mathrm{s}\mu_n= \Omega_\mathrm{s,pl}+\Omega_\mathrm{s,curv}+N_\mathrm{s}\mu_n,
\end{align}
where $\sigma_\mathrm{s}$, $\sigma_\mathrm{c}$ are the plane surface and the curvature tensions respectively, and $\mu_n$ is the neutron chemical potential. The functions $g_\mathrm{s}(u)$ and $g_\mathrm{c}(u)$ are listed in Table~\ref{table}\footnote{The values $g_{\mathrm{c}}(u)$ differ by factor of 2 from Ref.~\cite{Nakazato_ea11}, which is likely absorbed there in the curvature tension when defining the curvature through the mean value.} and plotted in Fig.~\ref{fig:gs_gc_wc}. They represent the normalized surface area and the curvature correction and depend on the particular shape.

Considering only the leading surface term  $\Omega_\mathrm{s,pl}$, the authors of Ref.~\cite{Hashimoto+84} showed that the normalized surface area of spherical clusters can be reduced beyond some filling fraction by changing shape, thus leading to the traditional pasta sequence. This sequence consists of clusters with threshold filling fraction $u_{\rm t,cl}$ and their inverted configurations (holes) with transitions at 
symmetric filling fraction $u_{\rm t,h}=1-u_{\rm t,cl}$. Here, however, as in Ref.~\cite{Nakazato_ea11} we take into account the curvature correction. From Table~\ref{table}, we see that $g_c(u>\pi/48)$ is lower for spaghetti than for spherical clusters. Therefore, spaghetti may appear for lower filling fractions when curvature correction is included, and similarly for other transitions (see Fig.~\ref{fig:gs_gc_wc}). The symmetry of $u_{\rm t}$ between clusters and holes is then broken. 

Finally, the Coulomb term can be factorized as \cite{Oyamatsu_ea84}
\begin{align}\label{eq:Coul}
     E_\mathrm{Coul}= (e y_p n_\mathrm{i})^2 w(u) a^5.
\end{align}
The function $w(u)$ is specified in Table~\ref{table} within the WS cell approximation and displayed in Fig.~\ref{fig:gs_gc_wc}. For the sake of comparison, its values for the body-centered cubic (bcc) lattice of spherical clusters/bubbles and hexagonal (hex) lattice of spaghetti/bucatini are also drawn.  
Note, that in these cases, thanks to large 
filling fraction,
the Coulomb energy can be calculated via summation over the reciprocal lattice, with no need to involve the Ewald approach (see, e.g., Ref.~\cite{Oyamatsu_ea84}).

\begin{table}[h]
\caption{Functions $g_\mathrm{s}(u)$, $g_\mathrm{c}(u)$ and  $w(u)$ corresponding to the surface, curvature and Coulomb terms (see Eqs.~\eqref{eq:surf}, \eqref{eq:Coul} respectively) for sphere, spaghetti and lasagna shapes. For bucatini (Swiss cheese) the functions $g_s(1-u)$, $w(1-u)$, $-g_c(1-u)$ corresponding to spaghetti (sphere) should be used.}\label{table}%
\begin{tabular}{@{}llll@{}}
\toprule
Shape & $g_s(u)$  & $g_c(u)$ & $w(u)$ \\
\midrule
sphere    & $(6\pi^{1/2}u)^{2/3}$   & $2(48 \pi^2 u)^{1/3}$ & $(9\pi)^{1/3}(2u^{5/3}-3u^2+u^{8/3})/(5\cdot2^{1/3})$  \\
spaghetti    & $(4\pi u)^{1/2}$   & $2\pi$ & $u^2(u-1-\ln{u})/2$ \\
lasagna     & 2   & 0  & $\pi u^2(1-u)^2/6$\\
\botrule
\end{tabular}
\end{table}

\subsection{Shape instability}\label{sec32}

In this section, we generally 
follow the line of arguments in Refs.~\cite{Hashimoto+84,Oyamatsu_ea84,Nakazato_ea11}. Namely, at given $\bar{n}$ and $u$, we compare the equilibrium energy density $E_\mathrm{CLD}^A/a^3$ of shape A  with the energy density of shape B ($E_\mathrm{CLD}^B/b^3$) keeping the same parameters $n_\mathrm{i}$, $n_{\mathrm{o}n}$, $n_{\mathrm{o}p}$ and $y_p$. 
This means that shape B has an energy density slightly above its equilibrium value. 
However, if $E_\mathrm{CLD}^B/b^3<E_\mathrm{CLD}^A/a^3$, then shape A is unstable. 
Using this criterion, the bulk energy $\mathcal{E}_\mathrm{bulk}(n_\mathrm{i},y_p,n_{\mathrm{o}n},n_{\mathrm{o}p},u)$ drops out. 
This is also the case for the energy density $N_\mathrm{s}\mu_n(n_{\mathrm{o}n},n_{\mathrm{o}p})/a^3$ 
of adsorbed neutrons since we compare configurations at the same density
$\bar{n}=un_\mathrm{i}+(1-u)n_\mathrm{o}+N_\mathrm{s}/a^3$. 
We denote the sum of the remaining surface and Coulomb energy terms for shape A and 
B as $E^A$ and $E^B$ respectively.  

For the shape A, the equilibrium condition with respect to the cell size $a$ reads 
\begin{align}
     \Omega_\mathrm{s,pl} + 2\Omega_\mathrm{s,curv} = 2E_\mathrm{coul} \quad.
\end{align}
This is a generalization of the ``virial theorem'' derived in Ref.~\cite{BBP1971} including 
curvature correction. 

With this equation, we find 
\begin{align}\label{eq:lambda-general}
    \lambda(\zeta )\equiv \frac{E^B }{E^A \zeta^3}
    =\frac{\Omega^B_\mathrm{s,pl} +\Omega^B_\mathrm{s,curv}}{\zeta ^3 (\frac32\Omega^A_\mathrm{s,pl}+2\Omega^A_\mathrm{s,curv})}+\frac{E^B_\mathrm{coul}}{\zeta ^3 (3E^A_\mathrm{coul}-\Omega^A_\mathrm{s,curv})} \quad,
\end{align}
where $\zeta\equiv b/a$. Then from Eq.~\eqref{eq:Coul}, we have
\begin{align}
   \frac{E^B_\mathrm{coul}}{E^A_\mathrm{coul}}=\frac{\zeta ^5w^B}{w^A}   \quad,
\end{align}
and from Eq.~\eqref{eq:surf}
\begin{align}
   \frac{\Omega^B_\mathrm{s,pl}}{\Omega^A_\mathrm{s,pl}}=\frac{\zeta ^2 g_\mathrm{s}^B}{g_\mathrm{s}^A}   \quad.
\end{align}

Introducing $X_\sigma$ as
\begin{align}
   \frac{X^A_{\sigma}}{a}\equiv \frac{\Omega^A_\mathrm{s,curv}}{\Omega^A_\mathrm{s,pl}}=\frac{g^A_\mathrm{c} \sigma_\mathrm{c}}{g^A_\mathrm{s} \sigma_\mathrm{s} a}   \quad,
\end{align}
Eq.~\eqref{eq:lambda-general} can be expressed as 
\begin{align}\label{eq:lambda}
     \lambda(\zeta )=\frac{2g^B_\mathrm{s}(1+X^B_\sigma/\zeta a)}{3\zeta g^A_\mathrm{s}(1+4X^A_\sigma/3a)}+\frac{\zeta ^2w^B}{3 w^A(1-2X^A_\sigma/3a(1+2X^A_\sigma/a))} \quad.
\end{align}

By optimizing the cell size of shape B, we obtain the following equation for the $\zeta $ factor:

\begin{align}\label{eq:minbox}
     -\frac{g^B_\mathrm{s}(1+2X^B_\sigma/\zeta a)}{\zeta ^2 g^A_\mathrm{s}(1+4X^A_\sigma/3a)}+\frac{\zeta w^B}{ w^A(1-2X^A_\sigma/3a(1+2X^A_\sigma/a))} =0 \quad.
\end{align}

Without curvature, i.e. $X_\sigma=0$, the solution is 
\begin{align}
     \zeta _0=\left(\frac{g^B_\mathrm{s} w^A}{g^A_\mathrm{s} w^B}\right)^{1/3} \quad,
\end{align}
and 
\begin{align}\label{eq:lambda_0}
     \lambda_0(\zeta _0)=\left(\frac{g^B_\mathrm{s}}{ g^A_\mathrm{s}}\right)^{2/3}\left(\frac{w^B}{w^A}\right)^{1/3} \quad.
\end{align}
In this case, we arrive at the criterion of Ref.~\cite{Oyamatsu_ea84}: if $\lambda_0(\zeta _0)<1$ then shape A is unstable with respect to shape B. We checked that using this criterion with the Coulomb energies for the bcc and hex lattices for spherical and cylindrical shapes respectively, we reproduce the results of Ref.~\cite{Oyamatsu_ea84} with pasta transitions occurring at $u_\mathrm{t}=[0.19,0.35,0.65,0.81]$ for the traditional sequence. From Fig.~\ref{fig:gs_gc_wc} we observe that the WS approximation underestimates the exact Coulomb energy, and the deviations are larger for spheres and Swiss cheese than for spaghetti and bucatini. With this approximation, we obtain a slight shift in the threshold filling fractions $u_\mathrm{t}=[0.215,0.355,0.645,0.785]$.
The resulting transitions obtained with this simple criterion reproduce very well the ones from full CLDM numerical calculations from Ref.~\cite{Lim&Holt17}, where curvature and neutron skin corrections were neglected, and the ones from Ref.~\cite{Zemlyakov+22}, where only curvature effects were discarded. As predicted by Ref.~\cite{Hashimoto+84}, the $u_{\rm t}$ are universal, irrespective of the nuclear interaction. 

The refined criterion of Ref.~\cite{Nakazato_ea11} can be reproduced accounting for the curvature correction perturbatively up to the first order. 
Namely, remarking that $X_\sigma$ should be much smaller than 
$a$, we substitute $\zeta _0$ in the curvature term of Eq.~\eqref{eq:minbox} thus leading to the approximate solution
\begin{align}\label{eq:zeta_1}
\zeta_1=\left(\frac{g^B_\mathrm{s} w^A(1+2X^B_\sigma/\zeta _0 a)(1-2X^A_\sigma /3a(1+2X^A_\sigma/a))}{g^A_\mathrm{s} w^B(1+4X^A_\sigma /3a)}\right)^{1/3} \quad.
\end{align}
Hence, the instability condition becomes $\lambda(\zeta _1)<1$. 
Subsequently, expanding $\lambda(\zeta_1)$ to first order in $X_\sigma/a$, we recover the result of Ref.~\cite{Nakazato_ea11}. 
However, here we use Eqs.~\eqref{eq:lambda} and \eqref{eq:minbox} without any further approximation.\footnote{We also numerically checked that inserting Eq.~\eqref{eq:zeta_1} in Eq.~\eqref{eq:lambda} yields identical results at the given precision unlike the expansion adopted in Ref.~\cite{Nakazato_ea11}.}

One can notice that the inclusion of the curvature correction breaks the universality of the threshold filling fractions. Accounting for $X_\sigma/a$ inevitably shifts all the $u_{\rm t}$ to lower values, and the shift is not universal, i.e.\ it
depends on the applied nuclear functional since it is controlled by the ratio $\sigma_\mathrm{c}/\sigma_\mathrm{s}$. 

\subsection{Numerical results}\label{sec33}

In principle, the surface tensions employed in CLDMs can be calculated within the (E)TF approach for semi-infinite matter in plane geometry \cite{Kolehmainen_ea85,Ravenhal_ea_surfT_83,Centelles_ea98,DHM00} and then parametrized as a function of the proton fraction $y_p$ (or neutron excess) inside the denser phase. Instead, the surface tension is often fitted directly to reproduce experimental atomic masses~\cite{DinhThi+21a,DinhThi+21b,Parmar+22,Lim&Holt25}. However, this can lead to the spurious and uncontrolled inclusion of various nuclear-physical effects like shell corrections, which cannot be incorporated adequately within the CLDM. In contrast, this problem does not arise when fitting  $\sigma_\mathrm{s}$ and $\sigma_\mathrm{c}$ to the full ETF nuclear mass table, as was done in Ref.~\cite{Carreau+20}. Moreover, such a table spans the whole nuclear chart from proton to neutron drip lines and therefore incorporates more information on large isospin asymmetries. We make use of the parametrization of Ref.~\cite{Carreau+20} for the BSk22, BSk24, and BSk25 functionals. For comparison, we also apply the fit to the experimental atomic masses for BSk24 and SLy4 from Ref.~\cite{DinhThi+21a}. The parameterization form of Refs.~\cite{Carreau+20,DinhThi+21a} originally introduced in Ref.~\cite{Lorenz_PhD91} reads

\begin{align}\label{eq:sigma_c}
     \sigma_\mathrm{c}(y_p)=\sigma_\mathrm{s}(y_p)\alpha\frac{\sigma_\mathrm{0,c}}{\sigma_\mathrm{0,s}} (\beta-y_p) \quad,
\end{align}
where $\alpha$ is usually fixed to $5.5$. Therefore, the higher the values of $\sigma_\mathrm{0,c}$ and $\beta$ the more abundant are pasta in the crust. The correlation should be opposite with $\sigma_\mathrm{0,s}$. This observation is confirmed in the statistical analyses of Ref.~\cite{DinhThi+21a}. 

\begin{figure}[h]
\centering
\includegraphics[width=0.65\textwidth]{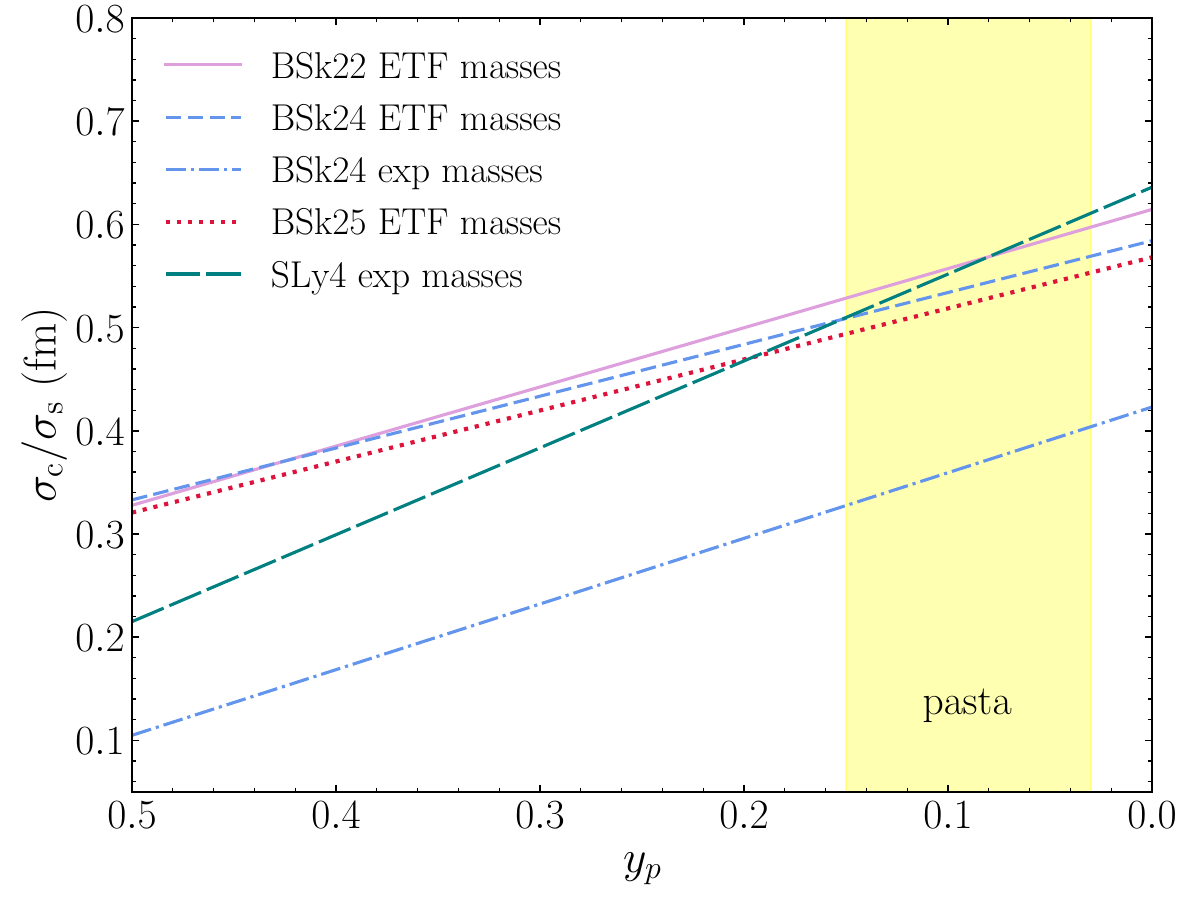}
\caption{The ratio of the curvature to the plane surface tension as a function of proton fraction in the dense phase $y_p$. The fit to the ETF nuclear mass tables from \cite{Carreau+20} and the fit to the experimental atomic masses from \cite{DinhThi+21a} are used. The yellow band indicates the $y_p$ interval corresponding to the pasta structures.}\label{fig:sigmas}
\end{figure}

We show the ratio $\sigma_\mathrm c/\sigma_\mathrm s$ in Fig.~\ref{fig:sigmas}. For BSk22, BSk24, and BSk25, the fit to the ETF nuclear masses leads to similar results over all proton fractions of neutron-rich clusters. Fitting experimental atomic masses instead, as for instance in Ref.~\cite{DinhThi+21a}, can lead to quantitatively different behavior. This is illustrated in Fig.~\ref{fig:sigmas} with BSk24. Such a comparison shows the importance of including in the fit information about systems at high isospin asymmetries. Nevertheless, for SLy4-based CLDM with surface terms fitted to experimental atomic masses, $\sigma_\mathrm c/\sigma_\mathrm s$ lies around $0.5$ fm - $0.6$ fm at $y_p\sim0.03-0.15$ (relevant for pasta) close to the values obtained for BSk22 - BSk25 CLDMs calibrated to ETF masses. 

\begin{figure}[h]
\centering
\includegraphics[width=0.6\textwidth]{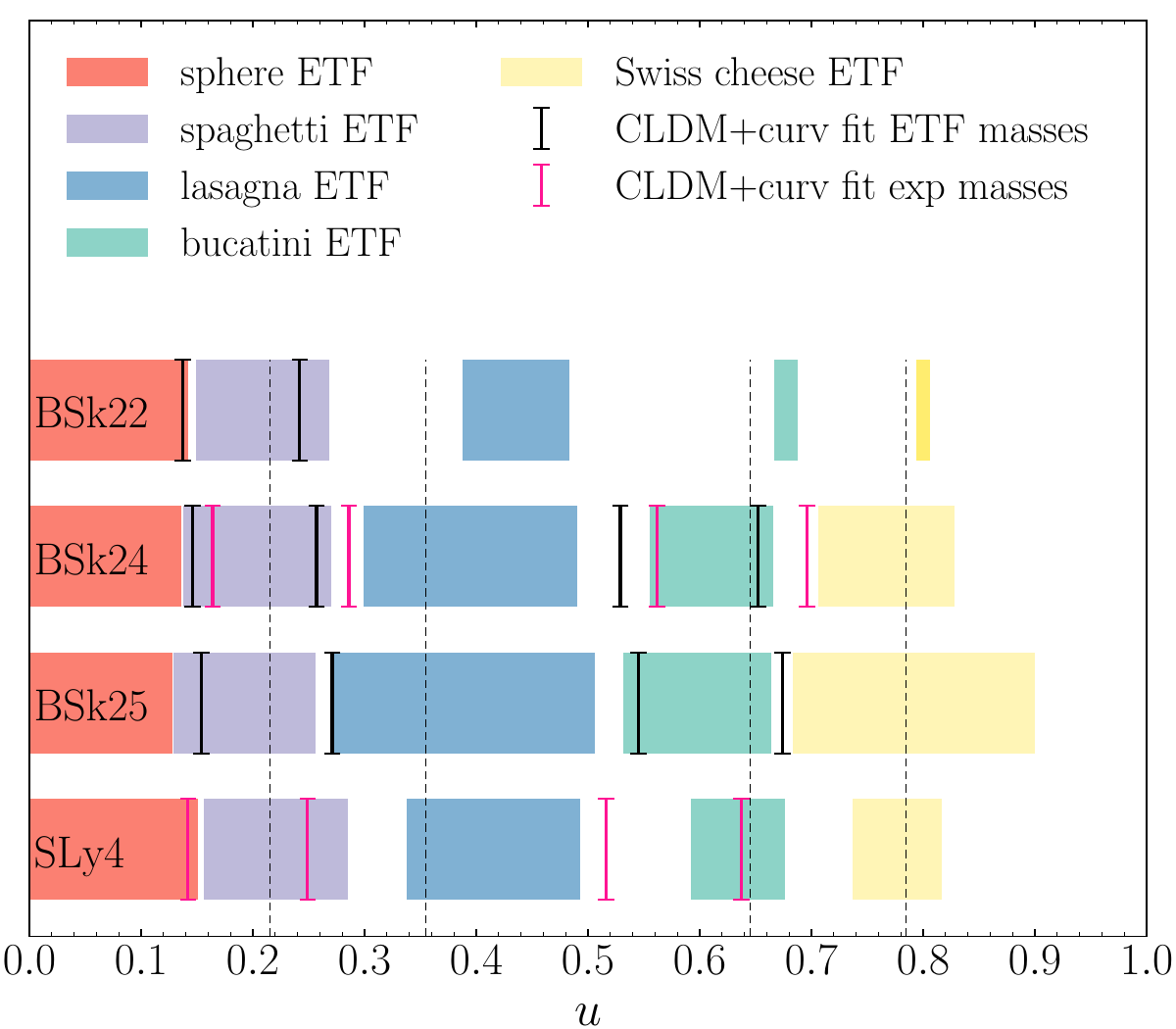}
\caption{Pasta sequences as a function of filling fraction $u$. Colored panels represent the results of the ETF calculations within different Skyrme functionals. The dotted lines indicate the pasta transition within the CLDM criterion of Ref.~\cite{Oyamatsu_ea84} without surface curvature correction. Black (pink) bars correspond to respective pasta transitions including the curvature effects using  Eqs.~\eqref{eq:lambda} and \eqref{eq:minbox} with curvature tension adjusted to the ETF \cite{Carreau+20} (experimental \cite{DinhThi+21a}) mass tables. See text for details.} \label{fig:pasta_sequence}
\end{figure}

To check the reliability of formulas~\eqref{eq:lambda} and~\eqref{eq:minbox}, we consider the BSk24 CLDM with $\sigma_\mathrm c$ fitted to experimental atomic masses and compare our predictions with the results of Ref.~\cite{DinhThi_PhD24} (note that due to the specific form of Eq.~\eqref{eq:sigma_c}, the parameters of $\sigma_{\rm s}$ are not required for the analysis). For this purpose, we use the size of the WS cell of shape A and its proton fraction as a function of $u$ as obtained from our ETF calculations. The computed threshold filling fractions for the instabilities of spheres, spaghetti, lasagna and bucatini are respectively given by $u_\mathrm{t}\approx[0.16,0.29,0.56,0.7]$ and are 
indicated by pink vertical bars in 
Fig.~\ref{fig:pasta_sequence}. We find good agreement with the extensive CLDM calculations of Ref.~\cite{DinhThi_PhD24} thus proving the correctness of the criterion $\lambda(\zeta)<1$ using Eqs.~\eqref{eq:lambda} and \eqref{eq:minbox}.

To assess the impact of the fitting protocol of the surface terms in the CLDMs, we now compare the predictions of the instability analysis using $\sigma_\mathrm c$ calibrated to ETF masses for BSk24. As can be seen by the vertical black bars in Fig.~\ref{fig:pasta_sequence}, the corresponding threshold filling fractions are closer to the full ETF results. Turning to the SLy4 CLDM with $\sigma_c$ fitted to experimental masses, we notice that at least the first transition is predicted in good agreement with our ETF calculations carried out for this functional in the same manner as described in Sec.~\ref{sec2}.

We see that the filling fraction for the onset of pasta inferred from the criterion $\lambda(\zeta)<1$ using Eqs.~\eqref{eq:lambda} and \eqref{eq:minbox}
is not very sensitive to the adopted functional and resides around $u_\mathrm{sp}\approx 0.14-0.15$, very close to the range of our ETF predictions. Such a good agreement is only achieved when curvature corrections are included.
This confirms the general expectations discussed in Sec.~\ref{sec32} that it is this correction that shifts the onset of pasta to lower values. For the remaining transitions, the agreement between the instability condition and the ETF results deteriorates due to (i) the limitations of the CLDM description of the smoothly varying nucleon density profiles, (ii) the lack of data on extremely neutron-rich systems beyond neutron drip in the fit of $\sigma_\mathrm{c}$, and (iii) discontinuities in $u$ in ETF calculations (see Sec.~\ref{sec2}). For example, for  BSk22, the lasagna phase is found to remain stable with respect to the criterion $\lambda(\zeta)<1$ up to the crust-core transition.
This is why only the first two shape transformations within the CLDM are shown in Fig.~\ref{fig:pasta_sequence} for this functional. Nevertheless, the bucatini and Swiss cheese are found to exist with BSk22 in an extremely narrow range of baryon densities in the ETF approach, and can hardly be seen in Fig.~\ref{fig:n_sequence}. 

In all cases, taking into account curvature effects in the surface terms of the CLDM substantially improves the agreement with ETF results, and explains the quasi-universality of the threshold filling fractions $u_{\rm t}$. 

\section{Conclusions}\label{sec4}

In this paper, we study the formation of nuclear pasta phases in neutron stars within the second-order extended Thomas-Fermi approach adopting the precision-fitted functionals BSk22, BSk24, and BSk25 associated with different behaviors for the symmetry energy $S(n)$.
Using the more realistic nucleon density profile parametrizations introduced in Ref.~\cite{ShchechilinParam+23}, we extend and clarify the correlation between the existence of pasta and the symmetry energy at relevant densities, previously discussed in Ref.~\cite{Shchechilin+sym23}. Namely, for all models, we observe very similar threshold filling fractions $u_{\rm t}$ corresponding to the pasta shape transitions. This conclusion coincides with the prediction previously made within the simplest version of the CLDM~\cite{Hashimoto+84,Oyamatsu_ea84}. However, the ETF values of $u_{\rm t}$ are systematically lower. The symmetry energy is then found to control the density dependence of $u(\bar n)$ and therefore determines the density ranges of the various pasta configurations (the lower the symmetry energy, the stiffer $u(\bar n)$  therefore the narrower the density range for each pasta).

To understand the quasi-universality of $u_{\rm t}$ and their values, we have further analyzed the 
stability of the different pasta shapes within the CLDM generalizing the arguments of 
Refs.~\cite{Hashimoto+84,Oyamatsu_ea84,Nakazato_ea11} by considering both curvature and neutron-skin corrections. Our results confirm that the inclusion 
of curvature effects leads to a shift of all threshold filling fractions to lower values, in agreement with 
Refs.~\cite{Nakazato_ea11,Newton_ea13_Survey}. In particular, the onset of pasta changes from $u_\mathrm{sp}\approx0.215$ 
to $u_\mathrm{sp}\approx0.14-0.15$ in a similar range with our ETF calculations using the same functionals. The exact shift depends on the ratio of the curvature to the plane surface tension $\sigma_{\rm c}/\sigma_{\rm s}$: the higher 
this ratio, the lower $u_\mathrm{sp}$. However, the good agreement between the CLDM and ETF calculations requires 
accurate calibration of the surface tension to extremely neutron-rich clusters in dense environments, as found 
in neutron-star crusts. 

For comparison, we also considered the SLy4 functional used in the popular CLDM of Ref.~\cite{DH00,DH01} with the curvature correction of Ref.~\cite{DHM00} and for which no pasta were found despite $u$ reaching about 0.28 at the crust-core
transition. Repeating our analysis for this functional, we obtain $u_\mathrm{sp}\approx0.14$ close to the 
values for the BSk22, BSk24, and BSk25 functionals. Moreover, the existence of pasta we find for SLy4 
is consistent with the CLDM calculations of Ref.~\cite{DinhThi+21a}. 

The simplified instability criterion $\lambda(\zeta)<1$ with Eqs.~\eqref{eq:lambda} and \eqref{eq:minbox} can be used to estimate the abundance of pasta in neutron stars without detailed numerical calculations of the various pasta configurations. This criterion can also provide a valuable 
guide in the search for the optimum pasta shape within (E)TF approaches. 
However, the plane surface and curvature tensions remain very uncertain at the extreme isospin asymmetries prevailing in the deepest layers of neutron-star crusts.
Moreover microscopic 
corrections such as proton shell effects can significantly alter the equilibrium pasta shapes  \cite{Pearson+22,Shchechilin+sym23,ShchechilinParam+23}.

\backmatter

\bmhead{Acknowledgements}

The work of N.N.S. was 
financially supported by the FWO (Belgium) and the Fonds de la Recherche Scientifique (Belgium) under the Excellence of Science (EOS) programme (project No. 40007501). The work of N.C. received funding from the Fonds de la Recherche Scientifique (Belgium) under Grant No. IISN 4.4502.19. The work of A.I.C. was funded by the baseline project FFUG-2024-0002 of the Ioffe Institute.

\section*{Declarations}

\begin{itemize}
\item Data availability This manuscript has no associated data or the data will not be deposited [Authors’ comment: The data described
in this manuscript are available from N.N.S. upon reasonable request.]
\item Code availability This manuscript has no associated code/software. [Author’s comment: Code/Software sharing not applicable to this article.]

\end{itemize}




\end{document}